\RequirePackage{desc-tex/docswitch}

\documentclass[12pt, twocolumn, apj]{openjournal}
\usepackage{desc-tex/lsstdesc_macros}
\usepackage{graphicx}
\usepackage{hyperref}
\usepackage[T1]{fontenc}
\usepackage{ae, aecompl}
\usepackage[utf8]{inputenc}
\usepackage[english]{babel}
\usepackage{amsmath,amstext}
\usepackage{fontawesome}
\usepackage{color, colortbl}
\usepackage{savesym}
\savesymbol{tablenum}
\restoresymbol{SIX}{tablenum}
\usepackage{booktabs, tabularx, multirow, bbm}
\usepackage{xcolor}
\usepackage{fancyvrb}
\usepackage[hang]{subfigure}
\graphicspath{{./}{../figures/}}

\usepackage{adjustbox}

\newbox\subfigbox 
\makeatletter
    \newenvironment{subfloat}
    {\def\caption##1{\gdef\subcapsave{\relax##1}}%
        \let\subcapsave=\@empty 
        \let\sf@oldlabel=\label 
        \def\label##1{\xdef\sublabsave{\noexpand\label{##1}}}%
        \let\sublabsave\relax
        \setbox\subfigbox\hbox
            \bgroup}%
            {\egroup
        \let\label=\sf@oldlabel
        \subfigure[\subcapsave]{\box\subfigbox}}%
\makeatother

\VerbatimFootnotes

\lefthead{Madireddy et al.}
\righthead{A Modular Deep Learning Pipeline for Galaxy-Scale Strong Gravitational Lens Detection and Modeling}

\begin{document}
\title{A Modular Deep Learning Pipeline for Galaxy-Scale Strong Gravitational Lens Detection and Modeling}
\author{Sandeep Madireddy$^*$$^{1}$}
\author{Nesar Ramachandra$^{2,3}$}
\author{Nan Li$^{4}$}
\author{James Butler$^{5}$}
\author{Prasanna Balaprakash$^{1}$}
\author{Salman Habib$^{2,3}$}
\author{Katrin Heitmann$^{3}$}
\author{The LSST Dark Energy Science Collaboration}

\email{$^*$smadireddy@anl.gov}

\affiliation{$^{1}$ Mathematics and Computer Science Division, Argonne National Laboratory,Lemont, IL, USA}
\affiliation{$^{2}$ Computational Science Division, Argonne National Laboratory, Lemont, IL, USA}
\affiliation {$^{3}$ High Energy Physics Division, Argonne National Laboratory,Lemont, IL, USA}
\affiliation {$^{4}$ National Astronomical Observatories of China, Beijing, China}
\affiliation {$^{5}$ Department of Statistics, University of California, Berkeley, CA, USA}



\begin{abstract}
Upcoming large astronomical surveys are expected to capture an unprecedented number of strong gravitational lensing systems. Deep learning is emerging as a promising practical tool for the detection and quantification of these galaxy-scale image distortions. The absence of large quantities of representative data from current astronomical surveys motivates the development of a robust forward-modeling approach using synthetic lensing images. 
Using a mock sample of strong lenses created upon a state-of-the-art extragalactic catalogs, we train a modular deep learning pipeline for uncertainty-quantified detection and modeling with intermediate image processing components for denoising and deblending the lensing systems. We demonstrate a high degree of interpretability and controlled systematics due to domain-specific task modules trained with different stages of synthetic image generation. For lens detection and modeling, we obtain semantically meaningful latent spaces that separate classes of strong lens images and yield uncertainty estimates that explain the origin of misclassified images and provide probabilistic predictions for the lens parameters. Validation of the inference pipeline has been carried out using images from the Subaru telescope's Hyper Suprime-Cam camera, and LSST DESC simulated DC2 sky survey catalogues.
\end{abstract}

\section{Introduction}
Gravitational lensing refers to the deflection of light rays as they traverse a space curved by the presence of massive astrophysical objects. 
In the present era of precision cosmology, gravitational lensing has become a powerful probe in many areas of astrophysics and cosmology, from stellar to cosmological scales. 
Galaxy-galaxy strong lensing (GGSL) is a particular case of gravitational lensing in which the background source and the foreground lens are both galaxies, and the lensing system is sufficiently massive to distort images of sources into arcs or even closed arcs (``Einstein rings''), depending on the relative angular position of the two objects.
Since the discovery of the first GGSL system~\citep{Hewitt1988}, many valuable scientific applications have been realized, such as studying galaxy mass density profiles \citep{Sonnenfeld2015,Shu2016b,Kung2018}, detecting the galaxy substructure~\citep{Vegetti2014,Hezaveh2016,Bayer2018}, measuring cosmological parameters~\citep{Collett2014,Rana2017,Suyu2017}, investigating the nature of high redshift galaxies~\citep{Bayliss2017,Dye2018,Sharda2018}, and constraining the properties of self-interacting dark matter candidates~\citep{Shu2016,Gilman2017,Kummer2018}.

The capabilities of next-generation surveys such as the Rubin Observatory's LSST\footnote{\url{https://www.lsst.org/}},  Euclid\footnote{\url{https://www.euclid-ec.org/}} and Roman Space Telescope\footnote{\url{https://wfirst.gsfc.nasa.gov/}} 
will increase the number of known GGSLs by several orders of magnitude~\citep{Collett2015}.
The forthcoming enormous datasets necessitate an analysis of GGSLs using automated procedures that operate efficiently and robustly, relying on the high uniformity and quality of the datasets. 
A key consideration for the analysis of GGSLs is the ability to detect a relatively small number (few hundreds to thousands) of strongly lensed galaxies from millions, if not billions of target images. The detection scheme needs to ensure a sufficiently small number of false positives so that we identify high-quality positive identifications that can be followed-up with telescopic observation for confirmation and higher-resolution studies. Given the expense of follow-up campaigns, the focus should be on the identification of ``golden'' candidates with a very low false positive probability rather than obtaining a large number of borderline or marginally confident cases.

To this end, several algorithms have been developed to detect GGSLs in image data by recognizing arc-like features and the presence of Einstein rings \citep{Gavazzi2014,Joseph2014,Paraficz2016,Bom2017}. 
More recently, efforts to automate GGSL detection have turned to machine learning (ML), in particular deep learning (DL) algorithms, given their strong performance in image recognition tasks. The strong gravitational lens detection challenge \citep{Metcalf2019} demonstrated the relative success of applying various ML techniques for automated detection of GGSL systems \citep{Jacobs2017, Petrillo2017, Ostrovski2017, Bom2017, Hartley2017, Lanusse2018, Avestruz2019} compared to traditional feature extraction techniques. 
\cite{Hezaveh2017} and \cite{Pearson2019} have shown the feasibility and reliability of using DL to model strong lenses as an efficient alternative to traditional parametric methods. \cite{Perreault-Levasseur2017} presented details on the estimation of the posteriors of constrained lensing parameters using a DL method; \cite{Morningstar2018, Morningstar2019} demonstrated the possibility of using ML and DL techniques to reconstruct source galaxies in GGSLs. More recently, works by \cite{canameras2020holismokes, he2020deep, li2020new, Wagner-Carena2021, pearson2021strong, park2021large} have also adopted ML and DL techniques for a variety of tasks ranging from lens detection through binary classification and lens modeling with regression approaches. These approaches were trained primarily using synthetic catalogs generated with varying degrees of fidelity in the underlying physical models.

In this paper, we address the growing need for an automated analysis of GGSLs in two steps. First, to date, only hundreds of galaxy-galaxy strong lensing systems have been confirmed by both photometry and spectroscopy \citep[][]{Treu2010SLReview}, which is insufficient to train and evaluate deep learning models with millions of hyperparameters. However, limited observational data of galaxy-galaxy strong lensing systems cannot adequately cover the corresponding feature space. Consequently, it is necessary to expand the sample size, diversity, and complexity of strong lensing systems by adopting simulations. Therefore, we created a dataset of 120,000 simulated images (60,000 GGSLs and 60,000 non-GGSLs) using a catalog of GGSLs and a state-of-the-art synthetic catalog of galaxies (cosmoDC2; \citealt{CosmoDC2}) combined with the strong lensing simulation program PICS (Pipeline for Images of Cosmological Strong lensing)~\citep{PICS}, resulting in one of the largest mock lensing catalogs available, which includes light profiles of galaxies with two components, multiple bands of information based on semi-analytic models, and noise and PSF models derived by ray-tracing the photons from above the atmosphere through the optics and to the camera. \citep[][]{Connolly2010}. 

Second, we develop a modular deep learning pipeline for automated lens detection and modeling for GGSLs consisting of four modules: denoising, deblending, lens detection, and lens modeling. We adopt deep residual network~\cite{he2016deep} (ResNet)-based architectures to denoize the original pixelized images and remove lens light in the deblending module. Lens detection and modeling modules perform classification and regression, respectively, and are modeled with a variational information bottleneck (VIB) framework~\citep{alemi2017VIB} that we enhanced with normalizing flow~\cite{kobyzev2020normalizing}. The normalized flow-enhanced VIB provides a robust framework to enhance interpretability and generalizability using a combination of probabilistic learning and deep learning. This is crucial for real-world science applications such as the strong gravitational lensing considered in this work.

We present the synthetic catalog generation strategy in Section 2 and discuss the details of the proposed deep learning pipeline consisting of denoising, deblending,  lens detection, and lens modeling in Section 3. Following this, we will present the results and insights from the synthetic catalogue in Section 4. In Section 5, we will discuss the validation performance of the presented pipeline on astonomical observations from HSC and on the LSST-DESC simulated DC2 sky survey. Finally, in Section 6 we summarize our findings and discuss potential avenues for future research and extensions.

\section{Data Preparation}
\label{sec:data_prep}
Only a few hundred GGSLs, confirmed by both photometry and spectroscopy, are available \citep{Treu2010SLReview, hsc3_jaelani, warps2016, des_lens}. Considering this amount of data is insufficient to train and evaluate deep learning models to search and model many thousands of GGSLs. This difficulty is exacerbated by the need for training data for denoising and deblending. Thus, we created a synthetic GGSL dataset containing 120,000 simulated images (60,000 GGSLs and 60,000 non-GGSLs). Additionally, the diversity and complexity of the training set are critical for the applicability of the pipeline to real observations. Hence, we include a two-component light profile of galaxies, multiple-band information, and realistic models of noise and Point Spread Functions (PSF) to make the mock dataset realistic.

Specifically, we implement the simulations of GGSLs following seven steps: 1) create populations of lenses and sources according to the given statistical properties of GGSLs; 2) build mass and light models of foreground lenses; 3) calculate deflection fields of the lenses; 4) construct light profiles of background source galaxies; 5) run ray-tracing simulations to create strongly lensed images based on the deflection fields and light profile of sources; 6) stack the lensed images and the foreground images of lenses; and 7) add the telescope noise as well as the telescope's PSF blurring.

The populations of lenses and sources are built upon a catalog of strong lenses \citep{Collett2015} (hereafter, Collett15) and a state-of-the-art extragalactic catalog \citep{CosmoDC2}. Collett15 provides a mass model of Singular Isothermal Ellipsoids \citep[SIEs][]{Kormann1994} and a light model of the Sersic profile \citep{Sersic1968} for both lens and source galaxies, and cosmoDC2 provides a further detailed light profile for galaxies, containing bulges and disks described by the Sersic profile. To connect the mass profiles from Collett15 and light profiles from cosmoDC2, we cross-match the apparent magnitudes, axis ratios, position angles, and redshifts of the galaxies from Collett15 and CosmoDC2. Expressly, we first model a lens galaxy as a smooth Singular Isothermal Ellipsoid described completely by using lensing strength, axis ratio, and position angle $(Par_{0})$. To create a balanced dataset, we sample the three parameters following flat distributions \citep{Pearson2019}. Some basic information (such as light models and redshifts of lenses and sources, $Par_{1}$) of a GGSL can be obtained by matching the three parameters with Collett15, and then a set of detailed parameters $(Par_2)$ can be achieved by matching $Par_1$ with CosmoDC2. Repeating the above process, we generate a catalog of parameters to generate mock images with PICS \citep{PICS}. The distributions of the parameters and their origins are shown in Table~\ref{tab:parameter_distributions}.

\begin{table}
\caption{\label{tab:parameter_distributions} Distributions and origins of the input parameters in the synthetic dataset. $U(a, b)$ denotes a uniform distribution with bounds $a$ and $b$. $R(\mu > 5)$ stands for choosing positions randomly from the area in the source plane where the magnification of the lens is greater than $5$.}
\begin{center}
\renewcommand{\arraystretch}{1.0}
\begin{tabular}{ l l}
\hline
{\bf Parameter} & {\bf Distribution/Origin} \\
\hline\hline
{\bf Lens galaxy} \\
\hline \hline
{SIE model} \\
\hline
{Lens center $(^{\prime\prime})$} & $(0.0, 0.0)$ \\
Einstein radius $(^{\prime\prime})$ & $\theta_E \sim U(0.5, 3.0)$ \\ 
Axis ratio & $q_{\rm lens} \sim U(0.2, 1.0)$ \\ 
Orientation angle & $\phi_{\rm lens} \sim U(-\pi/2, \pi/2)$ \\ 
Lens redshift & Collett15 \\
Source redshift & Collett15 \\ 
\hline 
{Two-component Elliptical Sersic light} \\ 
\hline
Total Apparent Magnitudes in $[g,r,i]$ & Collett15 \\
Total Half-light radius & Collett15 \\
Total Axis ratio & Collett15 \\
Orientation angle & Collett15 \\
Bulge to Total Fraction & CosmoDC2\\
\hline
Half-light radius of the bulge & CosmoDC2 \\ 
Axis ratio of the bulge & CosmoDC2 \\
Orientation angle of the bulge & CosmoDC2 \\
Sersic index of the bulge & CosmoDC2\\
\hline
Half-light radius of the disk & CosmoDC2 \\ 
Axis ratio of the disk & CosmoDC2 \\
Orientation angle of the disk & CosmoDC2 \\
Sersic index of the disk & CosmoDC2 \\
\hline \hline
{\bf Environment} \\
\hline \hline
External shear modulus & CosmoDC2 \\
Orientation angle & CosmoDC2 \\
\hline \hline
{\bf Source galaxy}\\
\hline \hline
{Tow-component Elliptical Sersic light} \\ 
\hline
Source center for non-lenses $(\prime\prime)$ & $U(-7.5, 7.5)$ \\
Source center for lenses $(\prime\prime)$ & $R(\mu > 5)$ \\
Total Apparent Magnitudes in $[g,r,i]$ & Collett15 \\
Total Half-light radius & Collett15 \\ 
Total Axis ratio & Collett15 \\
Orientation angle & Collett15 \\
Bugle to Total Fraction & cosmoDC2\\
\hline
Half-light radius of bulge & CosmoDC2 \\ 
Axis ratio of bulge & CosmoDC2 \\
Orientation angle of bulge & CosmoDC2 \\
Sersic index of bulge & CosmoDC2\\
\hline
Half-light radius of disk & CosmoDC2 \\ 
Axis ratio of disk & CosmoDC2 \\
Orientation angle of disk & CosmoDC2 \\
Sersic index of disk & CosmoDC2 \\
\hline
\hline
\end{tabular}
\end{center}
\end{table}

The mass model of an individual lens galaxy, as adopted by Collett15, is taken to be a singular isothermal ellipsoid.  This model is not only analytically tractable, but also consistent with individual lens models and lens statistics on length scales relevant for strong lensing \citep{Koopmans2006, Gavazzi2007, Dye2008}. Accordingly, the deflection maps can be given by the parameters of the positions, velocity dispersions, axis ratios, position angles, and redshifts of lenses, as well as the redshifts of the source galaxies, namely $\{x_1, x_2, \sigma_v, q_l, \phi_l, z_l, z_s\}$. Since $\{x_1, x_2\}$ can be fixed to $\{0, 0\}$ by centering the cutouts on the lens galaxies and the lensing strength (that is, the Einstein radius) can be given by $\theta_E = 4 \pi (\sigma_v/c)^2 D(z_l, z_s)/D(z_s)$, the parameter array can be simplified to $\{\theta_E, q_l, \phi_l\}$. Here, $c$ is the speed of light and $D(z_l, z_s)$ and $D(z_s)$ are the angular diameter distances from the deflector to the source and from the observer to the source, respectively.

To generate the corresponding images of the lenses and sources in a given lensing system with $\{{\theta_E, q_l, \phi_l}\}$, we match them with Collett15 with the properties of $\{{\theta_E, q_l, \phi_l}\}$ to find the redshifts, effective radii, and apparent magnitudes in the [g, r, i]-bands of lens and source galaxies separately, namely $\{z_l, R^{\rm eff}_l, mag^{[g,r,i]}_l\}$ and $\{z_s, R^{\rm eff}_s,  mag^{[g,r,i]}_s\}$. By extracting information from cosmoDC2, we include morphological properties of bulges and disks of galaxies in Collett15. A process of cross-matching between $\{z, R^{\rm eff}, mag^{[g,r,i]}\}$ of the galaxies in Collett15 and cosmoDC2 is implemented to assign the properties of bulges and disks to the corresponding galaxies. Furthermore, the projected positions of the sources in the lensing system are randomly chosen in the area where the lensing magnifications are greater than $5$ in the source plane. Based on the matched information, we first generate strong lensing images without noise and PSF blurring, and the pixel size of the images is $(20/512) arcsec$, about one fourth of the pixel size ($0.18 arcsec$) of the final mock. Where $20 arcsec$ is the size of the stamp and $512$ is the number of pixels per side of the stamp.

Noise and PSF modeling are used to make the images realistic using the models of a ground-based telescope from \cite{Collett2015} and \cite{Connolly2010}. For mimicking the noise behavior of the 10-year stacked LSST data, we design a noise model mixed between read noise, which is Gaussian-like, and shot noise, which is Poisson-like noise that can be calculated from the flux in the pixelized images. The PSF model is also a Gaussian function with different full width at half maximum in the $[g,r,i]$-bands, i.e., $FWHM_{[g,r,i]} = [0.811012, 0.769839, 0.754780]\,\,arcsec$. Examples of the mock data are shown in Fig.~\ref{Fig:Inference_deblend}. 
The non-lensing systems are generated in the same way except that the strong lensing effects have been removed by setting the deflection angles to zero. 

The generated catalogue includes 60k GGSLs and 60k non-GGSLs. Of these, 54k GGSLs and 54k non-GGSLs are randomly sampled and used as training data, and the other 12k images are used as test data to quantify the generalization performance.

\section{Deep Learning Pipeline for Training and Inference}

\begin{figure*}[tp]
\centering
\includegraphics[width=\linewidth]{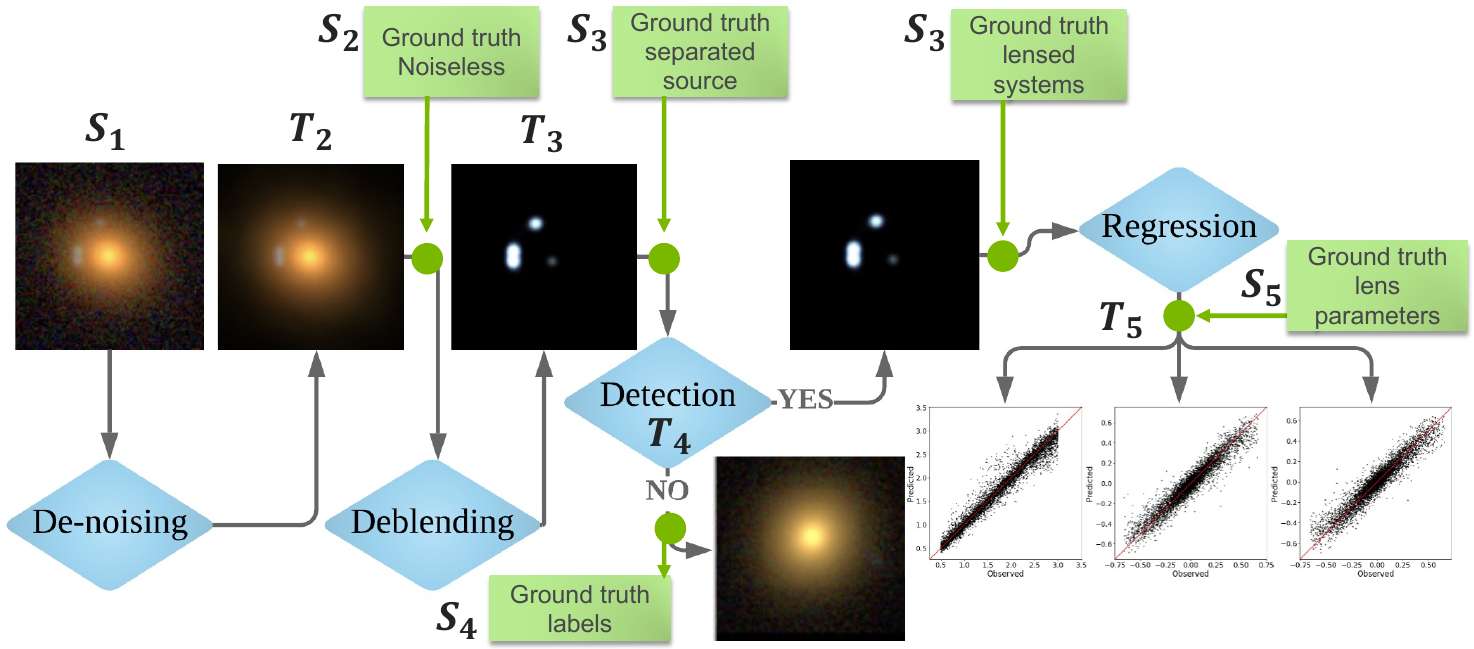}
\caption{Deep learning analysis pipeline for analysis of galaxy-scale strong lensed systems.}
\label{Fig:SL_pipeline_hor} 
\end{figure*}

The DL pipeline proposed in this work consists of four modules---denoising, source separation (deblending), lens detection (classification), and lens modeling (regression)---as shown in Fig.~\ref{Fig:SL_pipeline_hor}. A typical noisy observation from a telescope would be the input to this pipeline. The denoising module aims to remove the noise without affecting the background galaxy or the foreground lensed light. The denoised images are then passed through the deblending module to remove the foreground lensed light and output the image with just the background galaxy. This background galaxy image is then passed through the lens detection module, which determines whether an image contains a lensed or an unlensed galaxy. All galaxies labeled as lensed are then passed through the lens modeling module to identify their characteristics. The modular nature of the framework is necessary because each module is complex enough in itself to need separate validation. 
Moreover, having individual modules is critical for interpretability and provides flexibility to incorporate different hierarchies of domain knowledge into each module's model 
training without affecting other modules. We describe each of the four modules below. 

\subsection{Denoising and Deblending}
\label{sec:Denoise_deblend}
{Denoising} is an image restoration approach used to recover a clean image from
a noisy observation. 
Traditionally, image denoising has been posed as an inverse 
problem, where optimization approaches and special-purpose regularizers (known as image
priors) have been used~\citep{anwar2019deep}. Recently, deep learning-based approaches have been
increasingly adopted and are emerging as state-of-the-art algorithms~\citep{lim2017enhanced,zhang2018residual} for image denoising. 
We adopt a deep residual network-based enhanced deep super-resolution (EDSR)
architecture~\citep{lim2017enhanced}, which was developed for a specific type of image restoration. 
Since the inputs and outputs for denoising have the same resolution, we removed the upsampling layer from the EDSR architecture. This layer consisted of residual blocks $16$, each of which contained two convolutional layers and a nonlinear activation function $ReLU$. The convolution layers use $3 x 3$ kernels and $256$ feature channels.

The deblending module serves to decouple the lensed light and the source galaxy from the observations. This module adopts the same modified EDSR architecture as that used for the denoising module. The reason is that source separation is also an image-to-image task that takes the images with coupled source and the foreground galaxies as input and outputs the corresponding lensed or unlensed source galaxy that is separated from the foreground lens.

The denoising and deblending modules essentially preprocess the images in the pipeline to enhance the downstream lens searching and modeling tasks. The performance of these modules, especially when they are used for inference, is codependent: the output of one feeds as the input to the next module. We evaluated three different training strategies: 
(1) \emph{End-to-end training}, where a single model is trained so that its input is the noisy blended image and its output is the deblended image from the simulation data;
(2) \emph{Modular training}, where the denoising and deblending models are trained separately;  and
(3) \emph{Joint training}, where the denoising and deblending models are trained jointly by passing the output of the denoising model as input to the deblending model. Here, the loss function is a weighted combination of their individual losses.

The three strategies start with the same input. In modular and joint training, the input images to the deblending module were simulated noiseless blended images and predicted noiseless blended images, respectively. End-to-end training does not have this stage because it takes noisy blended images and directly outputs the noiseless deblended images. 
We use the L1 loss to train the denoising and deblending models, 
then employ the peak signal-to-noise ratio (PSNR) to evaluate the denoising and deblending accuracy; higher PSNR is better, with a maximum value typically close to 50 for 8bit images~\cite{sonka2014image}.

\subsection{Lens Detection}
\label{sec:lens_detection_pipe}
The lens detection module is a classification module that is used to detect lensing systems from source-separated images. In particular, each of the observed images needs to be classified as either a lensed or an unlensed system. We use the VIB approach to classification, which provides a unified framework for representation learning and predictive modeling, and compare its performance with the Resnet-50 model. 
We evaluated the classification model using the commonly used mean classification accuracy metric (mean over the two classes), the area under the receiver operating characteristic curve (AU-ROC), and the precision and recall metrics. The uncertainty in classifier predictions is quantified by using the entropy and conditional joint likelihood metrics~\citep{alemi2018uncertainty}.

\subsection{Lens modeling}
\label{sec:regression_theory}
The lens modeling module is a regression module that takes the source-separated galaxy and predicts its characteristics: Einstein radius, axis ratio, and position angle. We transform the axis ratio and position angle to two components of complex ellipticity because of the degeneracy between the axis ratio and position angle of a given lens. For example, combinations of $\{q, \phi\}$ and $\{1/q, \phi+\pi/2\}$ have the same morphology as an ellipsoid, leading to unreasonable scatters in modeling GGSLs with CNNs. Complex ellipticities $e_1$ and $e_2$ also avoid problems due to the periodicity of position angles.

The ability to quantify prediction uncertainty in addition to the point estimate is critical for validation purposes, in order to understand the impact of error sources (degraded quality of input data due to PSF, foreground objects, atmospheric conditions, detector noise) and modeling uncertainties (insufficient size, quality, and biases in the training samples and uncertainties associated with the network).  
We adopt and evaluate two strategies for lens modeling: VIB and a Resnet-101 model whose last dense layer is replaced by a stochastic denseFlipout~\cite{wen2018flip} layer (R101+DF).

\subsubsection{Variational Information Bottleneck} The VIB approach is a variational approximation of the information bottleneck principle (IB)~\citep{tishby2000information} proposed by~\cite{alemi2017VIB}. Taking into account the joint distribution $P(X,Y)$ of the input variable $X$  and the corresponding target variable $Y$, the IB principle defines an objective function that seeks to learn a latent encoding $Z$ that is maximally expressive about $Y$ while being maximally compressive about $X$. This can be formally written as a minimization problem:
\begin{equation}\label{eq:elbo}
    \underset{\theta}{\mathrm{min}}\,\,\mathcal{I}(Z,Y;\theta) - \beta\,\mathcal{I}(X,Z;\theta) = L ,
\end{equation}
where $\mathcal{I}(.,.;.)$ is mutual information and $\beta \geq 0 $ is the Lagrange multiplier that controls the trade-off between compressiveness and expressiveness of the model. 
For this reason,  several studies have shown that latent encoding using this approach leads to a better generalization capability~\citep{alemi2017VIB,achille2018emergence}.
The variational approximation of this objective can be written as follows.
 \begin{equation*}
 L \geq \frac{1}{N} \sum_{n=1}^{N} \int \mathrm{d}z p(z|x_n) \log q(y_n|z) - \beta p(z|x_n)\log\frac{p(z|x_n)}{r(z)},
\end{equation*}
where $p(z|x_n)$ can be modeled using an encoder and $q(y_n|z)$ by a decoder and $r(z)$ is the prior distribution in $z$, $x_n \in X$ is the $n^{th}$ input data. Following the standard convention of choosing the variational distribution of the latent variable to be a Gaussian distribution family and a standard Gaussian prior, we end up with the familiar objective function seen with standard variational autoencoders but with an additional $\beta$ factor. This objective can be maximized by using backpropagation using the reparameterization trick~\citep{kingma2013auto}:
 \begin{multline*}
 L \geq \frac{1}{N} \sum_{n=1}^{N} \mathbb{E}_{\epsilon\sim p(\epsilon)} -\log q(y_n|f(x_n,\epsilon)) \\ + \beta KL [p(Z|x_n), r(Z)].
\end{multline*}
where, $f(x_n,\epsilon)$ is the reparametrization and KL is the Kullback–Leibler divergence between distributions. We note that this approach combines representation learning and supervised learning in a single trainable framework and that the latent space can provide further insight into the model learning process to predict conditional distributions $P(Y|X)$, enhancing interpretability. In addition, VIB is a doubly stochastic approach in the sense that both the latent variable (Z) and the target variable (Y) are treated as random variables~\citep{alemi2018uncertainty} as opposed to deterministic approaches where only the latter is treated as a random variable. Therefore, this approach can be used to obtain uncertainty estimates of the predictions. 

For the lens detection task, the input variable $X$ corresponds to the source-separated galaxy represented as an image, while the target variable $Y$ is the binary class that indicates whether the image is of a lensed or unlensed galaxy. For the lens modeling task, the input variable $X$ corresponds to the source-separated galaxy represented as an image, while the target variable $Y$ corresponds to the three lens parameters (Einstein radius, axis ratio, and position angle). Details about the encoder and decoder architectures, as well as the latent space dimensions used in this work, are described in Section~\ref{Sec:Results_Regression_training}.

The conditional joint likelihood metric~\citep{alemi2018uncertainty} used to assess the uncertainty in the class label prediction for VIB is defined as 
\begin{equation*}
   p(Y,Z|X) = \int {d}X p(X)p(Z|X)q(Y|Z) \approx r(Z)q(Y|Z).
\end{equation*}
To quantify the uncertainty in lens detection VIB model predictions, the conditional joint likelihood metric and entropy are used. 

\subsection{Training and Inference Setup Details}

The simulation model used to generate the synthetic catalog  
provides five forms of data that were used to train the modules in the pipeline: noisy-blended galaxy images ($S_1$), which represent the noisy observations from the telescope; noiseless-blended galaxy images ($S_2$), which represent the scenario where the observed noise is perfectly removed from $S_1$; noiseless-deblended images ($S_3$), where the background source galaxy is perfectly separated from the foreground lens light; data labels ($S_4$), which indicate whether an image in $S_3$ is lensed or unlensed; and lens parameters ($S_5$), which describe the properties of lensed galaxies.  
Our goal is to train the denoising, deblending, lens detection, and lens modeling modules of the pipeline separately using different forms of simulation data (and to evaluate the corresponding test data splits, referred to as {\em component-wise inference modality}) and then use the trained models to predict lens parameters directly from noisy and blended galaxy images $S_1$ (referred to as {\em end-to-end inference modality}). 

To evaluate the pipeline, we split the 120K images into 108K (90$\%$) for training and use the remaining 12K images (10$\%$)  to evaluate both component-wise and end-to-end inference modalities. We refer to the predicted outputs from the pipeline during component-wise and end-to-end inference modality as $T_i \mid i \in \{2,3,4,5\}$ and $I_i \mid i \in \{2,3,4,5\}$, respectively, where $i$ represents the successive components in the pipeline. Separate terminology is required to differentiate the predicted outputs from the modules during component-wise and end-to-end inference. Some examples of these images are shown in Figure~\ref{Fig:Inference_deblend}. Finally, we validated the inference pipeline using images from the HSC catalog, obtained from real astronomical observations.

We adapt an open-source implementation of EDSR \citep{lim2017enhanced} written in PyTorch~\citep{paszke2017automatic} for denoising and deblending. VIB was implemented for lens detection and lens modeling in PyTorch by modifying the open source beta-tcvae implementation. ResNet architectures for lens detection and modeling are implemented in TensorFlow~\citep{abadi2016tensorflow}. All models were trained on a single compute node with a NVIDIA Tesla V100 GPU. Note that we have not used any additional data augmentation while training these models.

\section{Results and Discussion}

In this section, we present the training and inference results for each of the modules.

\subsection{Noise reduction and source separation}

Models used in the joint training strategy
used the same set of hyperparameter values as in the original EDSR paper~\citep{lim2017enhanced}. The number of epochs was set to $200$. Since EDSR trains by randomly extracting patches from the full image, it can learn to perform denoising and deblending from a smaller dataset. We used a subset of $18,000$ images ($9,000$ lensed, $9,000$ unlensed),  randomly selected from the 108K training split to reduce the computational cost of training the model. We found that the joint training strategy (see Section~\ref{sec:Denoise_deblend}) achieves higher accuracy and better stability compared to modular and end-to-end training. Modular training with different denoising and deblending models led to artifacts in the deblending model due to a discrepancy in the prediction of the denoised model (albeit very small). However, the joint training strategy led the deblending model to become robust to this discrepancy. For the end-to-end training strategy,  the model quickly diverged and its prediction results were poor.
\begin{figure*}[!t]
\includegraphics[width=1\linewidth]{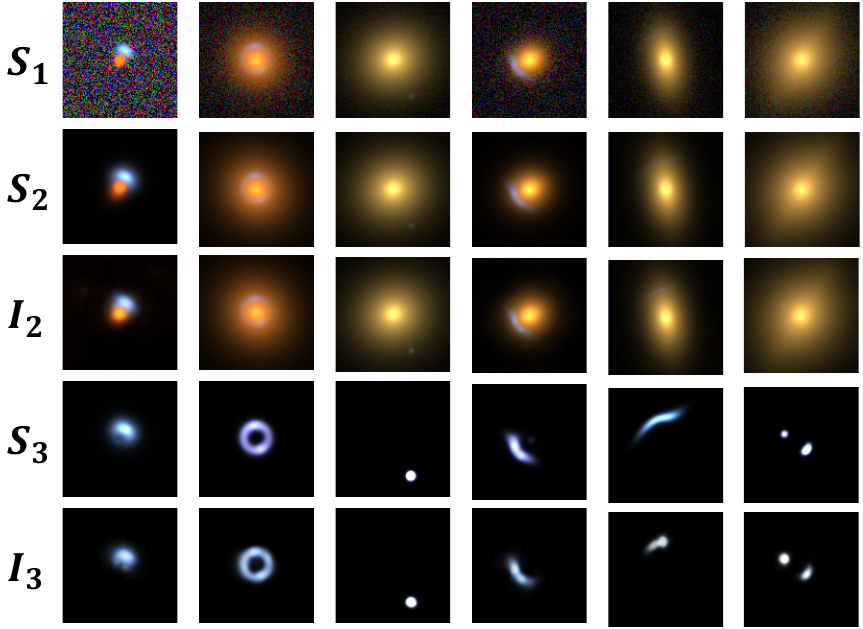} 
\caption{First row: noisy blended simulation data ($S_1$); Second row: noiseless blended simulation data ($S_2$); Third row: output from the denoising module at inference ($I_2$); Fourth row: noiseless deblended simulation data ($S_3$); Fifth row: output from the deblending model ($I_3$);} 
\label{Fig:Inference_deblend} 
\end{figure*}

The accuracy metrics for the joint training strategy are shown in Table~\ref{tab:Test_data_denoising}. In the case of denoising, we first calculate the PSNR value of the difference between $S_1$ and $S_2$, whose mean value in the test data split is $22.89$. The low value of PSNR demonstrates the significant difference between noisy and noiseless simulation data. This motivates the need to train a model to remove this noise. To this end, we compared the denoising model prediction ($T_2$)  with the same ground truth ($S_2$) in the test data split and found that the PSNR value was $45.66$. This demonstrates that the denoising model has learned to remove noise.

Similarly, for the deblending case, we compared $S_2$ with $S_3$, whose mean value in the test data split is seen to be $13.63$. The low PSNR value indicates a significant difference between the blended and deblended images, which is expected, since the latter contains only the background galaxy in the image. We also compared the deblending module prediction ($T_3$)  with the same ground truth ($S_3$) and found that the PSNR value in the test data split is $32.69$. This indicates a good recovery of the source galaxy by deblending. 

\begin{figure}[!htp]
\centering
\includegraphics[width=\linewidth]{"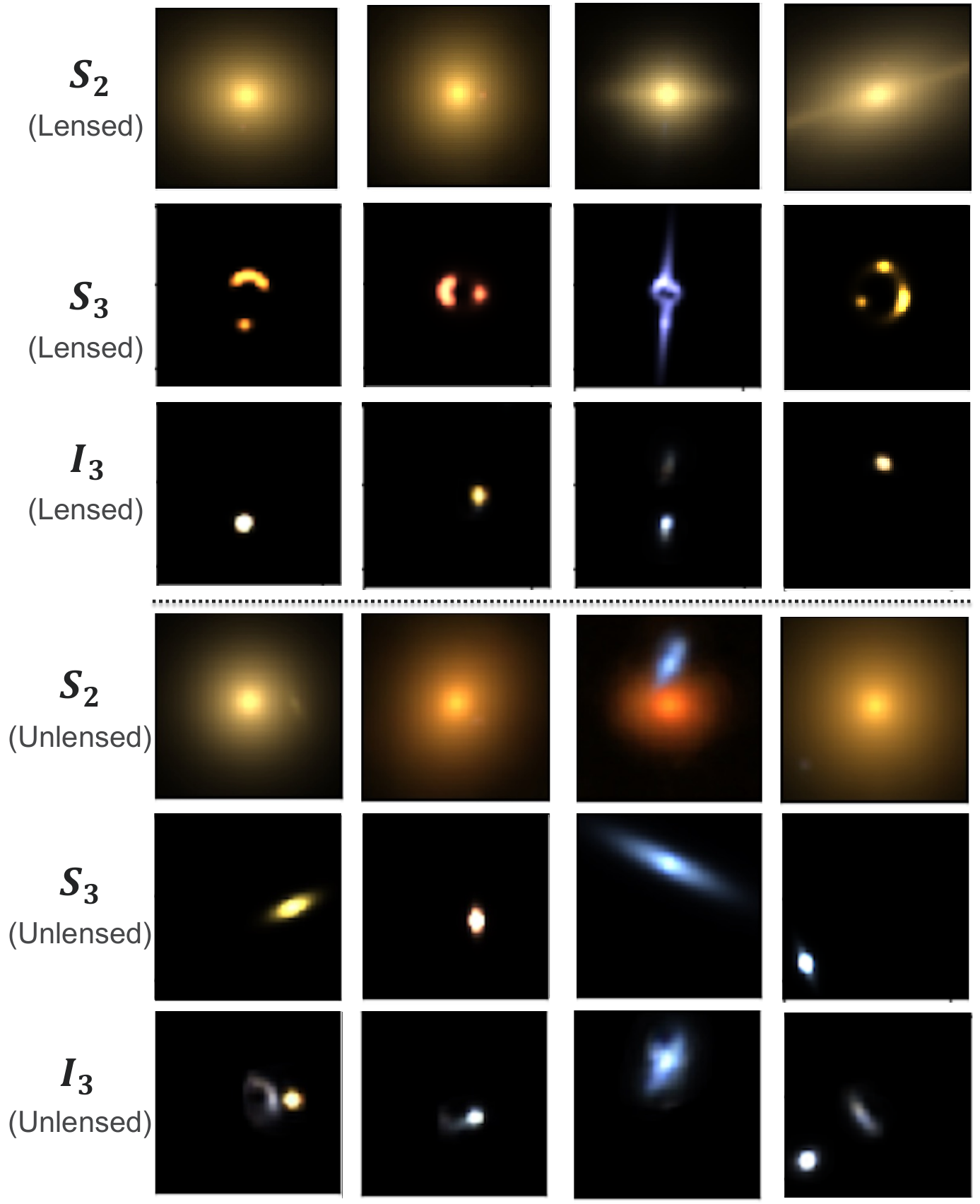"}
\caption{First row: noiseless deblended simulation data that are true lensed ($S_3$); second row: output from the deblending model ($I_3$) that are false negatives from classifier; Third row: noiseless deblended simulation data that are true unlensed ($S_3$); Fourth row: output from the deblending model ($I_3$) that are false positives from classifier.} 
\label{Fig:false_pred} 
\end{figure}

\subsection{Identification of strong lenses}
As discussed in Section~\ref{sec:lens_detection_pipe}, we adopted the VIB approach 
for lens detection. The encoder in the VIB model consists of $6$ convolutional layers with batch normalization between each of the layers, with the output of the last layer corresponding to twice the latent dimension (to model the mean and standard deviation) of the latent space, which is modeled as 10-dimensional mean field Gaussian distribution. We used factorial normalizing flow (FNF)~\cite{chen2018isolating} on top of this as a flexible prior, where each dimension is a normalizing flow of depth 32. The FNF can fit multimodal distributions and is therefore ideal for the classification module. The decoder, on the other hand, consists of four fully connected neural network layers that take a 10-dimensional input from the latent space to a Bernoulli distribution with a logit parameterization for the binary class prediction. The hyperparameters in the model closely follow those of ~\cite{chen2018isolating} (learning rate of $1e-3$ and batch size of 2048, as we adopted minibatch-weighted sampling from~\cite{chen2018isolating}) except that the number of epochs and $\beta$ terms from Equation~\ref{eq:elbo} were set at $300$ and $0.5$, respectively, based on a simple grid search on both these parameters. 

As a baseline, we trained a ResNet50 model (similar the model choice of ~\cite{lanusse2018cmu}) to predict the label directly from the noisy blended simulation images ($S_1$) and evaluated the metrics on the same test data. The model was trained with a batch size of 512 (the maximum batch size that we could fit into the GPU memory) and $150$ epochs (since we saw that the model converged and was not improving) with the rest of the hyperparameters matching the VIB model. Both models were trained on the 108K image training data and tested on the rest of the 12K image test data. The mean classification accuracy (over two classes) as well as AU-ROC, precision, and recall metrics on the test data were used to measure the accuracy of the classification model (Table~\ref{tab:Test_data_classification}). {In \em{component-wise inference modality,}} the mean accuracy was $0.82$. The model trained with $S_3$ as input gave a mean accuracy of $0.99$ with ResNet50 and $0.99$ with VIB and,  for the two models, AU-ROC of $0.99$ and precision and recall of $0.99$ and $0.99$, respectively, showing a significant improvement over the baseline. 
For {\emph{end-to-end inference modality}}, the lens detection module was fed the output of the deblending module $I_3$ as input to predict $I_4$. For the two models (ResNet50, VIB), we obtained a mean accuracy of ($0.93$, $0.94$), AU-ROC of ($0.96$, $0.97$), precision of ($0.93$, $0.94$) and recall of ($0.92$, $0.93$), respectively.

\begin{figure}%
\centering 
\subfigure[Latent space comparison for the test data split in component-wise (left) and end-to-end inference (right) modalities. The left (red) cluster corresponds to (true) lensed and (blue) right cluster corresponds to (true) unlensed. A clear separation exists between the two classes in the latent space for the former while the misclassified data are at the intersection of the clusters for the latter.]{    \includegraphics[width=0.475\linewidth]{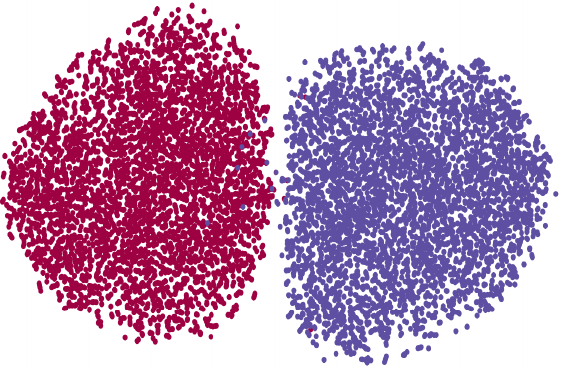}
    \hfill
    \includegraphics[width=0.475\linewidth]{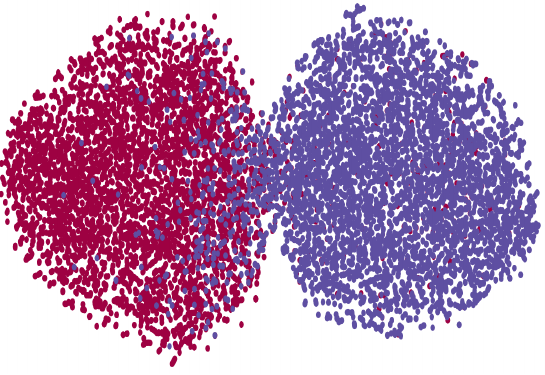}}\\ 
\subfigure[Conditional joint likelihood metric for end$-$to$-$end inference modality (colorbar on the right shows this metric value). The misclassified lensed data correspond to low metric value and hence high uncertainty.]{\includegraphics[width=\linewidth]{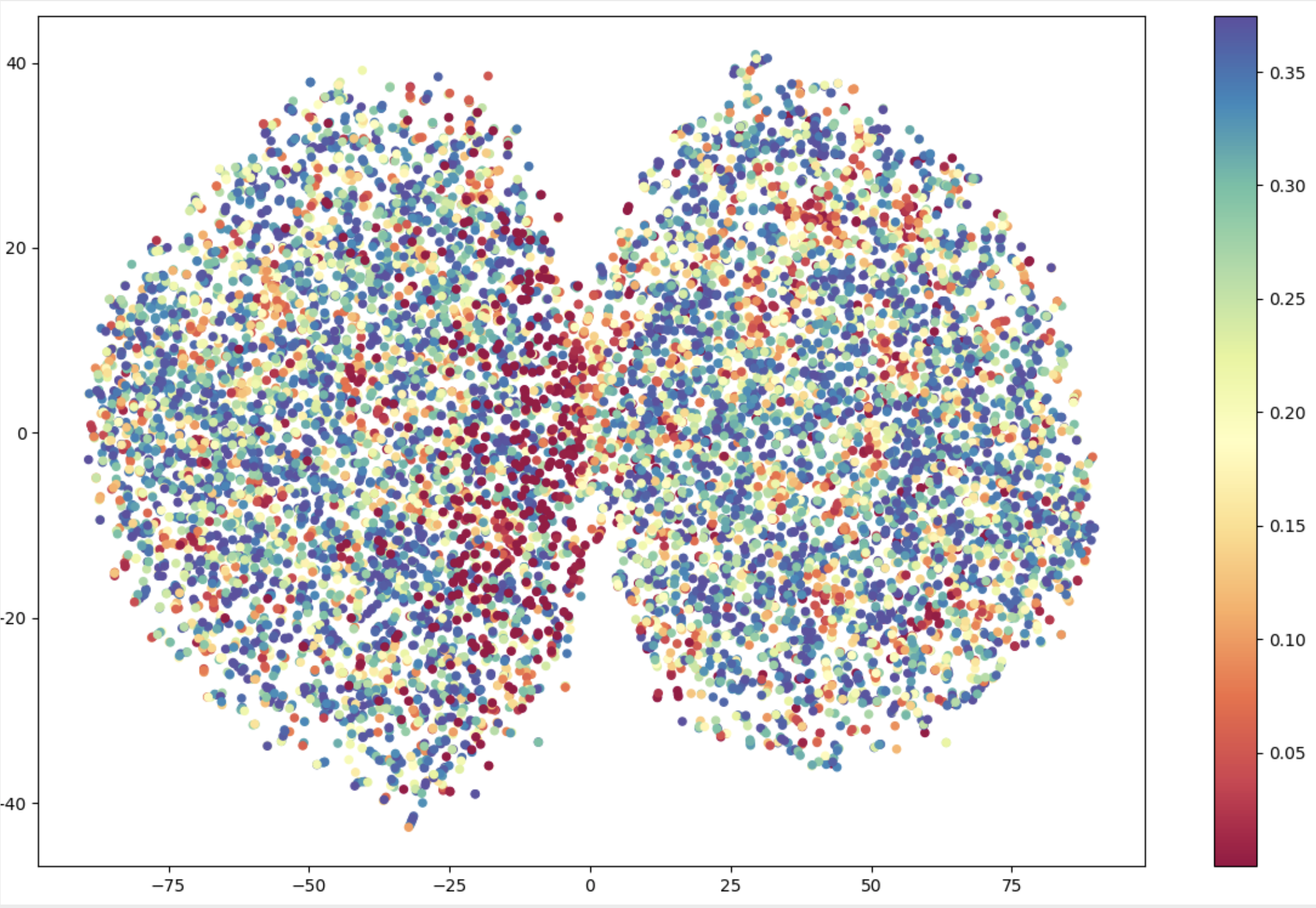}}%
\caption{Latent space for the lens detection VIB model.} \label{Fig:Classification_latent_output}
\end{figure}

The VIB model provides additional insights into the decision-making strategy by visualizing the latent space and calculating the uncertainty metrics of entropy and conditional joint likelihood. Figure~\ref{Fig:Classification_latent_output}(a) shows the two dimensional T-SNE projection of the latent space for the 12k test data split of $S_3$ images used as input to the model on top and $I_3$ input on bottom. There is a clear separation between the two classes in the latent space for $S_3$; with $I_3$ as inputs, there is no clear separation, but a small overlap, since the class assignments are sensitive to the learned denoising and deblending models. In addition, we note that the misclassified ones are at the intersection of the modes. To determine the reason for this misclassification, we calculate the conditional joint likelihood metric for the \emph{end-to-end inference modality} (Figure~\ref{Fig:Classification_latent_output}(b)) and found that for the left cluster (that corresponds to the true lensed data), the images falsely classified as lensed correspond, in fact, to low values of conditional joint likelihood, showing a higher uncertainty in the model predictions. Further inspection revealed that these points correspond to low magnification and low signal-to-noise ratios, which are difficult for the deblending model to predict. The entropy metric shows that the model is more uncertain about the images right at the intersection of the two classes. A few false positive and false negatives from the inference pipeline are illustrated in Fig. \ref{Fig:false_pred}.


\begin{table}[!t]
    \caption{Accuracy metrics for component-wise and end-to-end inference. $S_1$,$S_2$,$S_3$: Noisy blended, noiseless blended, and noiseless deblended simulation data; $S_4$: Data labels that indicate whether $S_3$ is lensed or unlensed; $S_5$: Lens parameters that describe the properties of the lensed galaxies; $T_2$,$T_3$: output from the denoising and deblending module respectively in the component-wise inference modality; $T_4$, $I_4$: data labels predicted by the lens detection module in the component-wise and end-to-end inference modality respectively; $T_5$, $I_5$: Lens parameters predicted by the lens modeling module in the component-wise and end-to-end inference modality respectively.}
    
    \begin{subfloat}
    \centering
        \caption{ PSNR metrics for denoising and deblending on $12000$ test split images (the component-wise and end-to-end modality is same in this case because we adopted the joint training stategy for denoising and deblending). }
        \label{tab:Test_data_denoising}
            \begin{adjustbox}{width=0.42\textwidth}
            \begin{tabular}{|c|c|}
            \hline
             {\it Component-wise/end-to-end inference} & \textbf{PSNR} \\ \hline
            \textbf{Denoising} \\ 
            \hline
            $S_1$ vs. $S_2$  & 22.89 \\ \hline
            $S_2$ vs. $T_2$  & 45.66 \\ \hline
            \textbf{Deblending} \\ 
            \hline
            $S_2$ vs. $S_3$  & 13.63\\ \hline
            $S_3$ vs. $T_3$  & 32.69  \\ \hline
            \end{tabular} 
            \end{adjustbox}
    \end{subfloat}%

  \begin{subfloat}
        \centering
        \caption{Component-wise and end-to-end Inference - Classification metrics on $12000$ test split images}
        \label{tab:Test_data_classification}
            \begin{adjustbox}{max width=0.45\textwidth}
            \begin{tabular}{|c|c|c|c|}
            \hline
             {\it Component-wise inference}&\textbf{Mean acc} & \textbf{AU-ROC}& \textbf{Prec, Rec} \\ \hline
            $T_4$(using $S_1$) vs. $S_4$ (ResNet50)  & 0.82  & 0.81 & 0.78 , 0.74   \\ \hline 
            $T_4$ vs. $S_4$ (ResNet50) & 0.99   & 0.99 & 0.99 , 0.99   \\ \hline
            $T_4$ vs. $S_4$ (VIB) & 0.99   & 0.99 & 0.99 , 0.99  \\ \hline
             {\it End-to-end inference}&\textbf{Mean acc} & \textbf{AU-ROC}& \textbf{Prec, Rec}  \\ \hline
            $I_4$ vs. $S_4$ (ResNet50) & 0.93 & 0.96 & 0.93, 0.92  \\ \hline
            $I_4$ vs. $S_4$ (VIB) & 0.94  & 0.97 & 0.94, 0.93  \\ \hline
            \end{tabular} 
            \end{adjustbox}
    \end{subfloat}
    ~
    \begin{subfloat}
        \caption{Component-wise and end-to-end inference - accuracy metrics on $6000$ lensed subset of test split images}
        \label{tab:regression}
             \begin{adjustbox}{max width=0.45\textwidth}
                \begin{tabular}{|c|c|c|}
                \hline
                {\it Baseline} & \textbf{MAE}-ResNet101 &  \\ \hline
                $T_5$(using $S_1$) vs. $S_5$ & 0.08 &  - \\ \hline
                 {\it Component-wise inference} & \textbf{MAE}-VIB &\textbf{MAE}-R101+DF  \\ \hline
                $T_5$ vs. $S_5$ & 0.01 &   0.09  \\ \hline
                 {\it End-to-end inference}& \textbf{MAE}-VIB &\textbf{MAE}-R101+DF\\ \hline
                $I_5$ vs. $S_5$ & 0.06 & 0.11 \\ \hline
                
                \end{tabular}
            \end{adjustbox}
    \end{subfloat}
\end{table}

\subsection{Estimation of lens parameters}

\label{Sec:Results_Regression_training}
For lens modeling, we again used the VIB approach. Only the lensed images in the deblended simulation data $S_3$ were used to train the regression model. Therefore, a total of $54,000$ was used for training and $6,000$  for testing. 
The encoder and decoder in the VIB model for lens modeling were similar to those used for lens detection, except that the decoder outputs a three-dimensional vector to predict the lens parameters with the likelihood chosen to be a Gaussian distribution. Similarly to the VIB lens detection model, the hyperparameters closely follow those in ~\cite{chen2018isolating}, except that the number of epochs and terms $\beta$ were set at $300$ and $3$, respectively, based on an empirical study. A thorough hyperparameter search will be considered in future work.

\begin{figure}[!tp]
\centering
\includegraphics[width=\linewidth]{"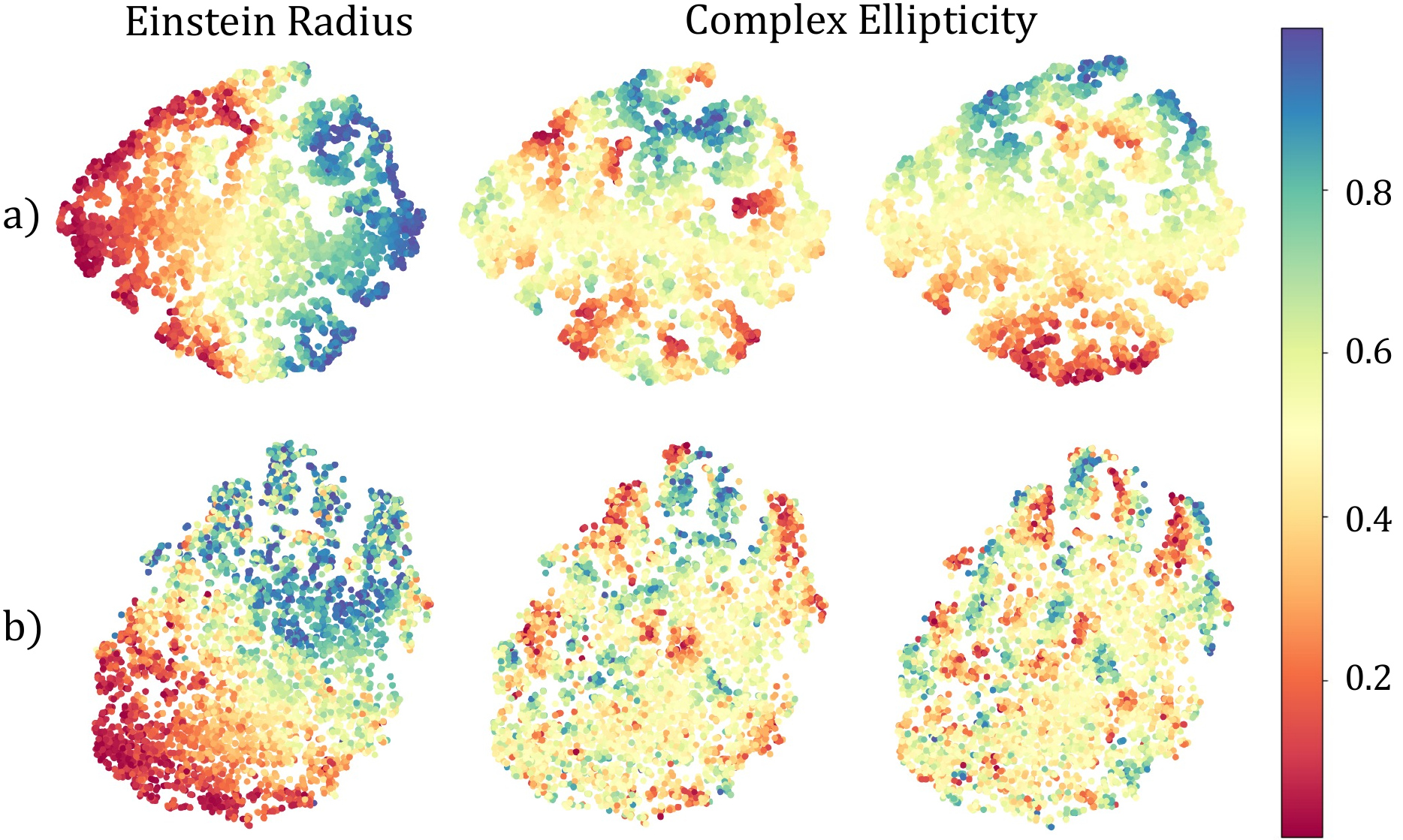"}
\caption{Visualization of the latent space (a) with the corresponding regression outputs (scaled to [0,1]) of the VIB model and (b) reconstruction of the input images.}
\label{Fig:Regression_latent_output} 
\end{figure}

\begin{figure*}[t!]
\centering
\includegraphics[width=0.9\linewidth]{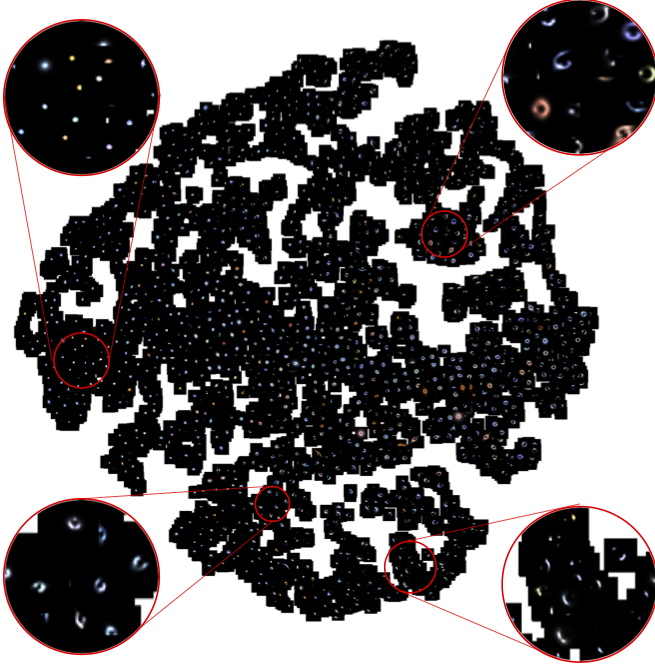}
\caption{Visualization of the latent space with the corresponding input images in VIB regression. Four zoomed-in panels show qualitatively different types of strong lenses.}
\label{Fig:Regression_tsne_images} 
\end{figure*}

For comparison, we used an additional architecture for the regression module, where we implemented a Resnet-101 model for parameter estimation, but with a densely connected layer class with Flipout estimator \citep{wen2018flipout} which is a variational inference approach that approximated the models weights (kernels and bias) with Gaussian distributions to transform it into a stochastic one. Furthermore, Flipout estimator improves the variational inference by decorrelating the gradients within a mini-batch by implicitly sampling pseudo-independent weight perturbations which helps in better variance reduction and hence faster and more stable training.
As a baseline, we trained a deterministic ResNet101 model to predict lens parameters directly from noisy blended simulation images ($S_1$) (similar to~\cite{Pearson2019}), and evaluated metrics on the same test data split. All of them were trained with the same hyperparameters as VIB except that the batch size was $512$ (maximum batch size given the memory of the GPU).

\begin{figure}[!tp]
\centering
\includegraphics[width=\linewidth]{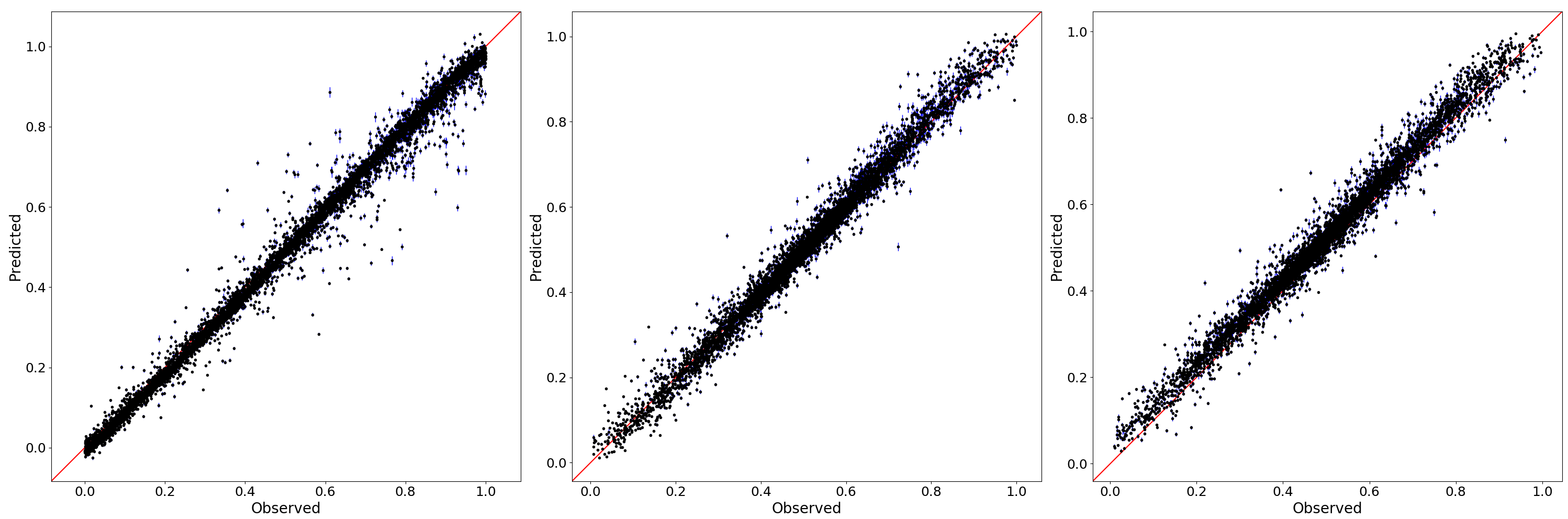}
\caption{Comparison of the observed ($S_5$) and predicted ($T_5$) data (and the error bars of one standard deviation) corresponding to the testing data for lens modeling (regression) in the training modality using VIB.}
\label{Fig:Regression_train} 
\end{figure}

The regression accuracy was measured by using the mean absolute error (MAE) in normalized coordinates ([0,1] with respect to the maximum and minimum of the training data), as shown in Table~\ref{tab:regression}. 
The plots comparing the observed and predicted data are shown in Fig.~\ref{Fig:Regression_train}. 
The MAE on the test data split for {\em component-wise inference modality} using the VIB model was $0.01$, which indicates a very good agreement with the ground truth. The corresponding accuracy for ResNet-101 with the Flipout estimator was much lower (MAE = $0.09$). This result indicates a superior performance with the VIB model. The baseline model trained on the $S_1$ data also had a higher MAE ($0.08$) in the test data split.

To gain additional insight into the VIB model, we visualized its latent space. Figure~\ref{Fig:Regression_tsne_images} shows a 2D t-SNE projection of the $10$-dimensional latent space where every sample of the latent space is labeled with the corresponding input image. The trend in the data can be analyzed by visualizing the characteristics of the images close together in this projected space. We can see that similarly shaped lenses are grouped together with the solid dots on the far left; these gradually change to hollow circles as we move right, and they change to arcs as we move to the bottom. To perform a quantitative analysis, we colored the latent samples with the value of the target variable (three in our case) as seen in Fig.~\ref{Fig:Regression_latent_output}(a). The results show that the projection of the latent space admits a gradual change in the Einstein radius from left to right and a complex ellipticity from bottom to top. The ability to obtain a good latent representation amenable to the regression task is a particularly important feature of the VIB model, enabling the model to obtain good regression accuracy, albeit with a shallower network and less training time. In Fig.~\ref{Fig:Regression_latent_output}(b) we show a similar 2D projection of the 10-dimensional latent space obtained when we reconstruct the input image instead of predicting the lens parameters (a common exercise for representation learning using variational auto encoders). We see that although the gradual change in the Einstein radius is admitted in the latent space, the other two variables do not have a trend. These results highlight an important point about the need to custom train the latent space for a given task in order to make the best use of the model. 

For the {\em end-to-end inference modality}, we found that the regression accuracy in the $6000$ images is slightly lower (MAE of $0.06$) than the accuracies obtained with the component-wise inference modality due to the propagation of errors from previous modules in the pipeline, but {\em end-to-end inference modality} with VIB models outperform the baseline of $0.08$. The Resnet101 model with dense Flipout layer has a lower regression accuracy with MAE of 0.11.

\section{Validation with real and synthetic sky images}

\subsection{Tests on Hyper Suprime-Cam Data}

We use observational data from the Hyper Suprime-Cam (HSC)\footnote{\url{https://www.naoj.org/Projects/HSC/}}, a digital camera on the Subaru telescope built by the National Astronomical Observatory of Japan. Recent gravitational lenses as part of the Survey of Gravitationally-lensed Objects in HSC Imaging (SuGOHI\footnote{Catalog of gravitational lenses from the SuGOHI lenses are publicly available here: \url{http://www-utap.phys.s.u-tokyo.ac.jp/~oguri/sugohi/}}) are identified by a combination of an arc-finding algorithm \textbf{YATTALENS} and visual inspection \citep{hsc1_sonnenfeld}. 
\vspace{0.1in}
\begin{figure}[!tp]
\centering
\includegraphics[width=\linewidth]{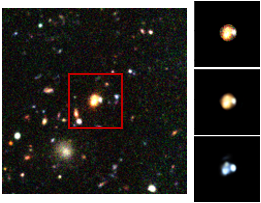}
\caption{Pre-processing of HSC observation by extracting the central ($111,111$) patch (left panel) and then applying a blob detection algorithm to extract the primary source in the center. The masked image (top right) is subsequently passed through the inference pipeline for denoising (middle right) and deblending (bottom right). }
\label{Fig:HSC_Process} 
\end{figure}
The default images from the HSC catalog are centered around the primary source, similar to our synthetic catalog. In an effort to minimize the contamination of unlensed images of galaxies on the line-of-sight, we cropped it to extract a central $(111,111)$ dimensional patch (which matches synthetic data), and then applied a blob detection algorithm~\cite{lowe2004distinctive} from Scikit-image~\cite{scikit-image} to find a circular mask that contains the primary source as seen in Fig.~\ref{Fig:HSC_Process}. Each of the preprocessed images is then passed through our inference pipeline to obtain the denoised, deblended images, as well as the classification labels. 

\begin{figure*}[!t]
\centering
\includegraphics[width=0.85\linewidth]{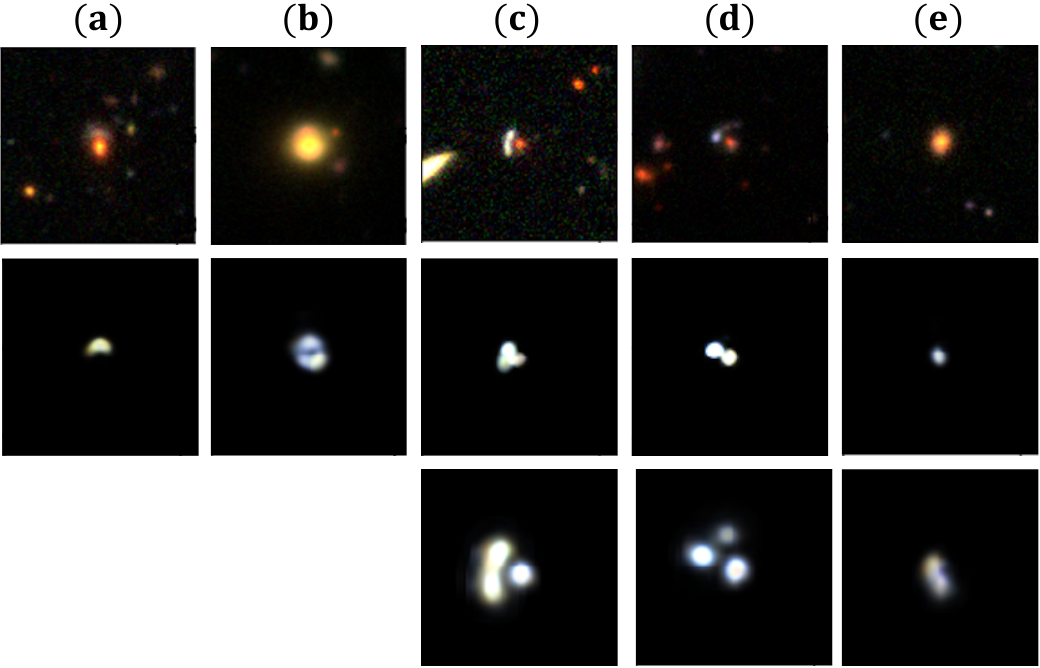}
\caption{The deblending output from the inference pipeline for preprocessed HSC images. (Row 1:) is the central $(111,111)$ patch, (Row 2:) is the corresponding deblended image, and (Row 3:) is the deblended output of the upscaled images. In (a,b), the images were correctly classified as lensed, while in (c,d) they were initially classified as unlensed but changed to lensed after upscaling. (e) Denotes the case where the upscaling did not affect the lens finding. }
\label{Fig:HSC_results} 
\end{figure*}

\subsubsection{HSC strong lens catalog}
Based on the confidence level of domain experts, the final catalog consists on galaxy-scaled strong lenses with categories grades A (definite lenses), B (probable lenses), and C (possible lenses), \citep{hsc2_wong}. Within this catalog, we `blindly' (i.e., without analyzing through the network) select a subset of fifty clean images for testing. Discarded samples include lensing systems with images of galaxies on the line-of-sight and off-centered images.
We found that 40 of the 50 images were correctly classified (for example, see Figure~\ref{Fig:HSC_results}(a,b)) as being lensed while the remaining were incorrectly classified (for example, see Figure~\ref{Fig:HSC_results}(c,d,e)). We note that all incorrectly classified images belonged only to class C. Further analysis of the incorrectly classified images revealed that the arc sizes 
of these images were too small compared to the synthetic catalog (see the typical arc sizes of the training sample shown in Figure \ref{Fig:Inference_deblend}). To mitigate this, we extracted a central $(50,50)$ patch from the wrongly classified images and upscaled it to increase the size of the arc. This led to a correct classification of seven of these images (for example, see Figure~\ref{Fig:HSC_results}(c,d)), thus bringing the classification accuracy to $94\%$. However, three ``wrongly'' classified images, although labeled lensed, looked ambiguous (for example, see Figure~\ref{Fig:HSC_results}(e)). The conditional joint likelihood metric, in fact, confirms this intuition by placing greater uncertainty on these observations. Furthermore, these three observations were also placed at one end of the (2D t-SNE projection of) latent space, separately from other images, further confirming the difference in features. This strategy of upscaling the images can be generalized in observational analyses, where testing images are scaled at multiple levels while forward propagating through Deep learning frameworks. Such a multi-scale approach not only increases the completeness in the strong lensing sample, but also provides a consistency check. Alternatively, the range of arc sizes in the training sample could also be enhanced via data augmentation strategies that are calibrated against survey specifications.

\subsubsection{Mix of lensed, unlensed and ambiguous cases}

Additionally, we collected data from the HSC catalog that were ambiguous for the human eye to classify into lensed or unlensed categories. All images in this dataset were manually assigned the probability of being lensed by a group of human observers. 
Images with probability less than $0.5$ are those that are ``most probably unlensed", while those with probability greater than $0.5$ are ``most probably lensed". 

All these images are passed through the inference pipeline, as previously stated. However, to gain further insight into why our pipeline fails to follow the human-labeled classes in few cases, we partition the misclassified images into the following five categories: (1) Missed capturing arc in the mask; (2) Extra images of unlensed galaxies in the light cone captured in the mask; (3) Correct mask but wrong prediction; (4) Correct mask but wrong prediction flipped to correct after zooming in; (5) Image is significantly different from training data (out of distribution). 
This particular categorization is intended to aid in the performance characterization of our mask extraction in the pipeline (to remove the images of line-of-sight galaxies) and to identify areas of potential improvement.

\begin{figure}[!htp]
\centering
\includegraphics[width=\linewidth]{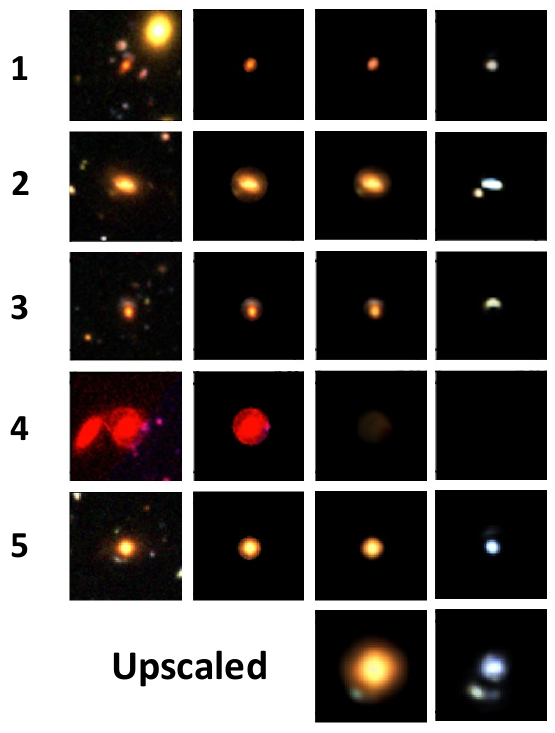}
\caption{Five categories in the ``most probably lensed" case (columns: Input, Mask, Denoised, and Deblended)}
\label{Fig:Five_categores_lensed} 
\end{figure}

\begin{figure}[!t]
\centering
\includegraphics[width=\linewidth]{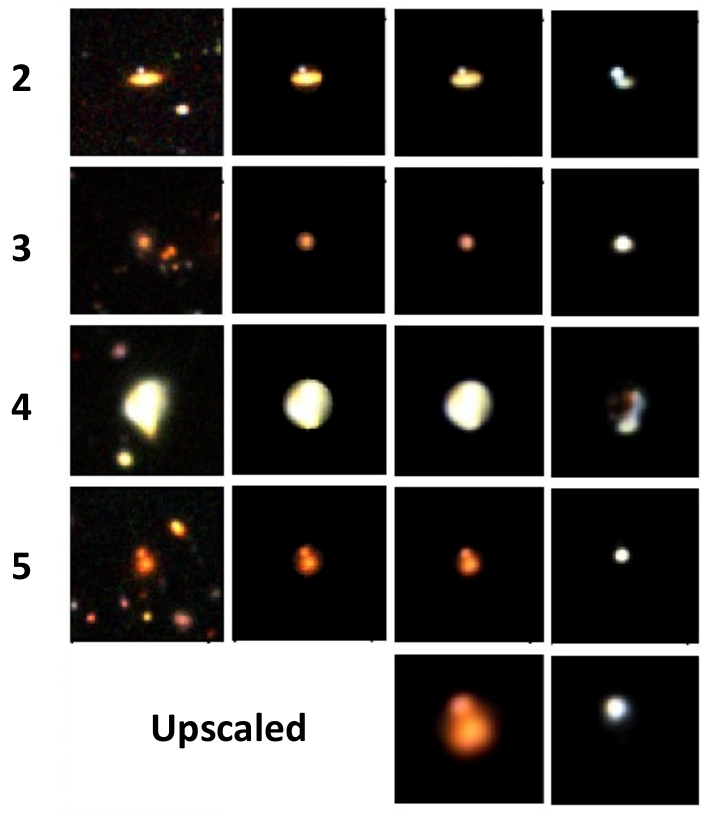}
\caption{Four categories in the case ``most probably unlensed" case (since category 1 doesnt exist for the unlensed case) (columns: Input, Mask, Denoised, and Deblended).}
\label{Fig:Five_categores_unlensed} 
\end{figure}

For the ``most probably lensed" cases (with a labeled probability greater than $0.5$), a total of $14$ out of $50$ were misclassified as an unlensed galaxy. Of $14$, there were $1, 3, 6, 3, 1$ images attributed to each of the five categories, respectively; representative examples of each of these five categories for this case are shown in Figure~\ref{Fig:Five_categores_lensed}. When we remove a small set of images where mask extraction was difficult (category $1,2,5$) and consider the images correctly classified after zooming in (category $4$), we have a total of $37$ of $43$ correctly classified, which is an accuracy rate of $86\%$. 
The same analysis for ``most probably unlensed" cases (with labeled probability less than $0.5$) found that a total of $16$ of $27$ were misclassified. A further categorization of these images showed $0, 5, 3, 4, 4$, respectively, in each category; representative examples from each of these five categories for this case are shown in Figure~\ref{Fig:Five_categores_unlensed}. A general observation highlighted by this subcategorization is that it is comparatively harder to separate unlensed images of galaxies on the line-of-sight and unlensed source galaxies in the unlensed mocks, resulting in more numbers ($5$ out of $16$) of images that capture line-of-sight galaxies in the mask. Additionally, we found that there are quite a few out-of-distribution examples ($4$). After removing the category $1,2,5$ we observe that $15$ of $18$ are correctly classified, leading to an accuracy of $83\%$.
The above findings highlight the difficulty in clean separation of secondary sources, which could potentially be overcome by utilizing data with secondary sources as part of the training set.

\subsection{Tests on DC2 synthetic sky data}

The vast majority of objects observed by LSST will be unlensed. Hence, it is important to benchmark a strong lensing pipeline against a large number of unlensed images, to quantify the number of false negatives (i.e., unlensed objects wrongly classified as lensed). This test is especially crucial for corner cases or the `lens imposters', like edge-on galaxies or disk galaxies. To validate our model against a large number of unlensed objects similar to LSST, we used simulated sky survey data, DC2 \citep{dc2, CosmoDC2}. 

We used the Run 2.2i DR6 Wide-Fast-Deep dataset (explained in \cite{dc2}) of the DC2 mock catalog that emulates 300 square degrees of LSST imaging over a 10-year campaign in 6 filters. The object data are extracted using the Generic Catalog Reader(GCR), a custom Python package originally developed as part of the DESCQA framework \citep{descqa}. The images are extracted from Butler (a middleware component for retrieving image datasets and executing pipelines for the Vera C. Rubin Observatory \citet{Butler2019, Butler2022}) in the g, r, and i bands. To match the GGSL input requirements, the postage stamps of 111x111 are created around the galaxy centers read from the GCR, and the band information is rescaled. 

\begin{figure}[!tp]
\centering
\includegraphics[width=\linewidth]{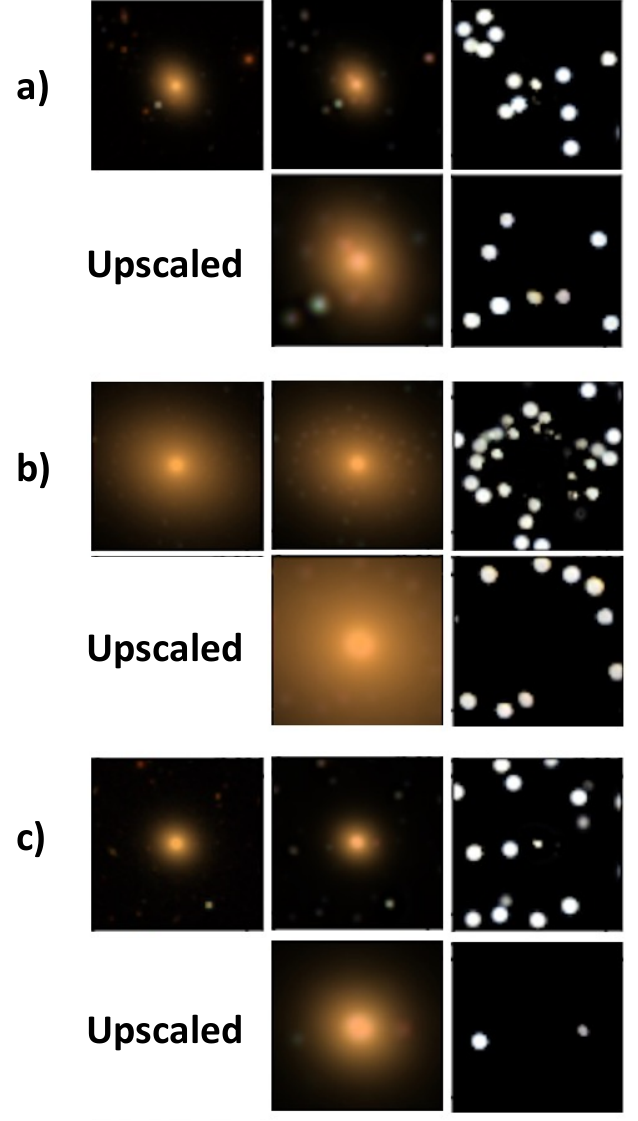}
\caption{Prediction of denoising and deblending in DC2 validation data (Columns: Input, Denoised, and Deblended). Each of a,b,c shows the denoised and deblended maps with and without upscaling. For a,b, we observed that the classifier incorrectly predicted that the snapshot was lensed in both the original and upscaled cases. For c, the classification label changes from lensed to unlensed after upscaling. Note that the DC2 catalogue does not contain lenses and is specifically used to study the false positive rates from our model. }
\label{Fig:DC2_Process_ambiguous} 
\end{figure}

In this particular catalogue, we observed significantly more unlensed images of galaxies on the line-of-sight and a more diffuse foreground source compared to the previous validation scenarios considered. This is primarily due to the design of the DC2 with up to magnitude depth of 28 in the LSST r-band. In addition, the combination of bulge and disk Sérsic profiles in DC2 synthetic sky may be too simplistic for our framework -- a shared issue with deblending and shape measurement pipelines in the LSST analysis pipelines \cite{dc2}. For this reason, a mask extracted with the blob detection algorithm (adopted for the previously mentioned datasets) intersected with the sources, and therefore we were unable to easily remove the line-of sight-images to a desired level. This led to poor lens-finding accuracy for this particular dataset. The extracted dataset, extracted, as well as the predicted models denoised and deblended in this dataset, are illustrated in Figure~\ref{Fig:DC2_Process_ambiguous}. In all the illustrated cases, we show the denoised and deblended maps with and without the upscaling operation. 
Similar to the HSC data, the effect of secondary sources seems to decrease the overall classification accuracy, and thus including them as part of the training data can further improve this accuracy

\section{Summary and Conclusions}
The combination of high-fidelity simulation data and a systematic machine learning pipeline is crucial to develop fast and accurate GGSL analysis techniques for future cosmological surveys. To this end, we generated a dataset of 120,000 synthetic images (60,000  GGSLs and 60,000 non-GGSLs), which we used to develop a deep learning pipeline with separate modules for denoising, deblending, lens detection, and modeling and validated it using real observations from the HSC catalog and DC2 catalogues. 
The modular nature of our pipeline allowed us to train, test, and evaluate each component in isolation and helped us understand its efficacy. 

We adopted the EDSR model for the denoising and deblending modules that provided a good recovery (PSNR of $45.66$ and $32.69$, respectively) of ground truth for both modules trained on the simulated data. For lens detection and lens modeling, we adopted the variational information bottleneck (VIB) approach and enhanced it with a normalizing flow that provided a superior accuracy (12\% improvement for lens detection and 25\% for lens modeling) over baseline and other deep ResNet architectures with only a small fraction of layers. The VIB approach also provides model interpretability by visualizing the latent space. The lens detection model produces a latent space that perfectly separates the two classes, thus demonstrating good representation learning ability when tested on the simulation inputs. With the inference pipeline, the higher-uncertainty images correspond to the misclassified images, which in turn belong to the low-magnification, low signal-to-noise ratio region, which is known to have difficulty in deblending. For lens modeling, we also find that the learned latent space with VIB has more semantic meaning compared to VAE. 

In order to validate our pipeline and evaluate it's accuracy and utility in real-world scenarios, we consider the observations from the Hyper Suprime-Cam (HSC) and LSST-DESC simulated DC2 sky survey catalogues. In the HSC case, we considered two different scenarios; the first is a lensed catalog, where the human observers categorized the observations to be strong lensed with high confidence. In the second scenario, we considered a catalog with observations that are classified by human observers with a higher level of ambiguity, thus we have the ``most probably lensed" and ``most probably unlensed" cases. Next, to validate against LSST-like observations, we generated a catalog of unlensed images by extracting cutouts from the LSST-DESC simulated DC2 sky survey. In the first scenario of HSC data, we found that our pipeline provided an accuracy of $94\%$, while in the second scenario for HSC, we observed an accuracy of $86\%$ and $83\%$ in the cases of ``most probably lensed" and ``most probably unlensed" after removing the cases where mask extraction was difficult. In the DC2 dataset, we observed significantly more {\bf line-of-sight galaxies} and a more diffused lens galaxy in general, so the images of line-of-sight galaxies were not effectively removed using the mask extraction adopted in the validation datasets previously used, leading to a poorer validation accuracy.

One limitation of this work is that relatively simple models of mass and light profiles of lens and source galaxies were used in the simulations that generated the training data. These choices lead to underestimated contamination from substructures in the context of both mass and light distributions of galaxies in the cutouts, in turn leading to either miscounted images or to introduce additional artifacts in the procedures of all modules. The issue is hardly noticeable in the data from ground-based surveys due to the large pixel size and spread of the PSF. However, it becomes significant in the data from space-based surveys with much higher spatial resolutions. To widen the scope of our pipeline, we plan to employ more realistic mass and light models, as well as to create a larger image dataset. Eventually, we intend to utilize the pipeline for real-time lens detection and modeling with data from next-generation large-scale sky surveys, including ground- and space-based telescopes such as Euclid, LSST, and Roman Space Telescope, where fast and automated methods of detecting and characterizing astronomical objects become a necessity.

The mask extraction was a primary part of the validation procedure because the data used to train the pipeline only consisted of the postage stamp data without the presence of substructures in the main lenses and secondary lenses on the line-of-sight, while the validation data from observations include the lensing effects from substructures and secondary lenses. Although the mask generation approach was successful in removing lens light in most cases, there were scenarios that the module can not deal with the perturbations on the lensed arcs due to the lensing effects beyond the smooth main lenses, such as the lensing systems involve significant lensing effects of substructures and secondary lenses. Similarly, the images of galaxies on the line-of-sight decrease the effectiveness of our mask extraction module, since they mimic multiply lensed images or/and bring blending effects in the scales smaller than the Einstein radius of the lens galaxy. To overcome these limitations, we have generated a synthetic catalog of lenses that includes substructures and secondary lenses. The light and mass of the substructures and secondary lenses are involved in the simulation, since the correlation between the light distribution of the substructures / secondary lenses and the lensing effects due to the substructures and secondary lenses would be helpful for high-quality extractions in the context of a low signal-to-noise ratio and small-scale perturbations for real observations. 
\newpage
\section*{Acknowledgments}

This material is based on work supported by the U.S. Department of Energy (DOE), Office of Science, Office of Advanced Scientific Computing Research, under Contract DE-AC02-06CH11357. It is also based on work supported by the U.S. Department of Energy, Office of Science, Office of Advanced Scientific Computing Research, and Office of High Energy Physics, Scientific Discovery through Advanced Computing (SciDAC) program. The authors gratefully acknowledge the computing resources provided and operated by the Joint Laboratory for System Evaluation (JLSE), the Argonne Leadership Computing Facility (ALCF), and the Laboratory Computing Resource Center (LCRC) at Argonne National Laboratory. The work is also supported by the UK Science and Technology Facilities Council (STFC).

The authors thank Anupreeta More for providing HSC data for the mix of lenses, unlensed, and ambiguous cases. The authors also thank Javier Sanchez for helping to extract DC2 postage stamps for synthetic data testing. 
This work has undergone internal review from the LSST Dark Energy Science Collaboration (LSST-DESC); we thank Camille Avestruz, Xiaosheng Huang, Ji Won Park and the LSST-DESC Publication board for the helpful comments during the internal review. 

\paragraph{Author Contribution Statement:} Sandeep Madireddy formulated the machine learning approach and developed the corresponding software, conducted the learning experiments and lead the manuscript writing. Nesar Ramachandra contributed to the data preparation, code development and manuscript writing. Nan Li developed the lensing simulation code, generated mock data, and contributed to manuscript writing. James Butler conducted experiments with the baseline machine learning approaches. Prasanna Balaprakash advised on the formulation of the machine learning pipeline and contributed to the manuscript writing. Salman Habib advised on the conceptualization of the developed approach as well as on analysis of model predictions both in training and validation experiments. Katrin Heitmann provided the underlying simulation for the paper, including specifically generated data products.

\bibliographystyle{apj}
\fontsize{9.0pt}{10.0pt} \selectfont
\bibliography{SL}


\end{document}